\documentclass[conference]{IEEEtran}
\IEEEoverridecommandlockouts
\usepackage[ruled,vlined,linesnumbered]{algorithm2e}
\usepackage{amsmath}
\usepackage{amssymb}
\usepackage{xcolor}
\usepackage{listings}
\usepackage{float}
\usepackage{framed}
\usepackage{fancybox}
\usepackage{graphicx}
\usepackage{multirow}
\usepackage{lipsum}
\usepackage{tabularx}
\usepackage{enumitem}
\usepackage{makecell}
\usepackage{soul}
\usepackage[most]{tcolorbox}
\usepackage{balance}
\usepackage{array}

\definecolor{mygray}{rgb}{0.98,0.98,0.98}

\makeatletter
\newcommand{\linebreakand}{%
  \end{@IEEEauthorhalign}
  \hfill\mbox{}\par
  \mbox{}\hfill\begin{@IEEEauthorhalign}
}
\makeatother

\usepackage{mdframed}
\mdfdefinestyle{customframe}{
    linecolor=black,           
    linewidth=1pt,            
    backgroundcolor=white, 
    roundcorner=10pt,         
    innerleftmargin=3pt,      
    innerrightmargin=3pt,     
    innertopmargin=3pt,       
    innerbottommargin=3pt,    
}

\AtBeginDocument{%
  \providecommand\BibTeX{{%
    \normalfont B\kern-0.5em{\scshape i\kern-0.25em b}\kern-0.8em\TeX}}}

\newtheorem{Definition}{Definition}

\newtheorem{Observation}{Observation}

\newcommand{\code}[1]{{\footnotesize\texttt{#1}}}

\newcommand{\tool}{\textsc{KAT}\xspace}

\newcolumntype{P}[1]{>{\centering\arraybackslash}m{#1}}

\begin{document}

\title{{\tool}: Dependency-aware Automated API Testing with Large Language Models}


\author{\IEEEauthorblockN{Tri Le}
\IEEEauthorblockA{\textit{Katalon Inc.} \\
Ho Chi Minh City, Vietnam \\
tri.qle@katalon.com}
\and
\IEEEauthorblockN{Thien Tran}
\IEEEauthorblockA{\textit{Katalon Inc.} \\
\textit{University of Science}\\
\textit{Vietnam National University}\\
Ho Chi Minh City, Vietnam \\
thien.tran@katalon.com}
\and
\IEEEauthorblockN{Duy Cao}
\IEEEauthorblockA{\textit{Katalon Inc.} \\
\textit{University of Science}\\
\textit{Vietnam National University}\\
Ho Chi Minh City, Vietnam \\
duy.cao@katalon.com}
\and
\IEEEauthorblockN{Vy Le}
\IEEEauthorblockA{\textit{Katalon Inc.} \\
Ho Chi Minh City, Vietnam \\
vy.le@katalon.com}
\and
\linebreakand
\IEEEauthorblockN{Tien N. Nguyen}
\IEEEauthorblockA{\textit{Computer Science Department} \\
\textit{University of Texas at Dallas}\\
Dallas, Texas, USA \\
tien.n.nguyen@utdallas.edu}
\and
\IEEEauthorblockN{Vu Nguyen\textsuperscript{*}}
\IEEEauthorblockA{\textit{Katalon Inc.} \\
\textit{University of Science}\\
\textit{Vietnam National University}\\
Ho Chi Minh City, Vietnam \\
nvu@fit.hcmus.edu.vn}
}

\maketitle
\thispagestyle{plain}
\pagestyle{plain}

\begin{abstract}
API testing has increasing demands for software companies. Prior API testing tools were aware of certain types of dependencies that needed to be concise between operations and parameters. However, their approaches, which are mostly done manually or using heuristic-based algorithms, have limitations due to the complexity of these dependencies. In this paper, we present {\tool} (Katalon API Testing), a novel AI-driven approach that leverages the large language model GPT in conjunction with advanced prompting techniques to autonomously generate test cases to validate RESTful APIs. Our comprehensive strategy encompasses various processes to construct an operation dependency graph from an OpenAPI specification and to generate test scripts, constraint validation scripts, test cases, and test data. Our evaluation of {\tool} using 12 real-world RESTful services shows that it can improve test coverage, detect more undocumented status codes, and reduce false positives in these services in comparison with a state-of-the-art automated test generation tool. These results indicate the effectiveness of using the large language model for generating test scripts and data for API testing. 
\end{abstract}


%


\maketitle
\let\thefootnote\relax\footnotetext{* Corresponding: Vu Nguyen (nvu@fit.hcmus.edu.vn)}

\section{Introduction}

RESTful APIs (or REST APIs - REpresentational State Transfer) is a software architectural style to guide the development of web APIs. RESTful APIs communicate through HTTP requests to perform standard database functions such as creating, reading, updating, and deleting records (CRUD). RESTful API has become one of the most used software architectural styles due to its high flexibility and scalability, as well as being relatively secure and easy to implement.

Automated test cases and data generation for API testing have been an active research topic that has attracted many studies in recent years \cite{viglianisi2020resttestgen, MartinLopez2021Restest, arcuri2019restful, liu2022morest, alonso2022arte, atlidakis2019restler}. These studies can be divided into black-box and white-box test generation approaches. Black-box testing, which is the most common, uses the OpenAPI Specification (OAS) as a basis to generate test cases and data. White-box testing focuses on analyzing source code to drive test cases and data \cite{arcuri2019restful}. 

While current approaches have made significant progress in testing RESTful APIs, they still face challenges in addressing {\em intricate dependencies among API endpoints and their parameters}. These dependencies fall into three categories. Firstly, there are dependencies among API endpoints (operations). For instance, to test an endpoint for charging a credit card in an online flight booking system, one must first invoke the operation for selecting the flight. Secondly, there are dependencies among the parameters of an operation. For example, in a flight booking system, the operation requires parameters like \code{arrivalDate} and \code{departureDate}. And in that context, the constraint is that \code{departureDate} must precede \code{arrivalDate}. Thirdly, there are dependencies between an operation and its parameters. For instance, in a flight booking system, the parameters \code{arrivalDate} and \code{departureDate} of a reservation operation must be set in the future. 

Unfortunately, the foregoing API testing frameworks have limitations in handling such dependencies. In the case of inter-operation dependencies, RestTestGen~\cite{viglianisi2020resttestgen} employs a heuristic approach centered on name matching to identify relationships among operations. This hinges on a shared field between the output of one operation and the input of another. However, discrepancies in field names could lead the heuristic algorithm to incorrectly establish dependencies. For instance, exclusive reliance on field names might fail to accurately identify the \code{GET /flights} endpoint as a dependent operation of the \code{POST /booking} endpoint, since the algorithm might not find matching pairs of field names.

In terms of inter-parameter dependencies, previous methods have indeed considered them in test case generation. Yet, it is crucial to note certain limitations in these approaches. For instance, RESTest~\cite{MartinLopez2021Restest} addresses this by mandating the manual inclusion of dependencies among parameters in the testing OAS file under the \code{x-dependencies} field. This manual intervention demands a significant amount of time from testers. Thus, any improvements in simplifying this process would signify progress in this field.

Regarding dependencies between operations and parameters, cutting-edge approaches employ heuristic-based methods (RestTestGen~\cite{viglianisi2020resttestgen}) or rule-based methods (bBOXRT~\cite{laranjeiro2021black}) to generate both valid and invalid test data. However, they might not fully comprehend these constraints articulated in natural language or might necessitate manual intervention.

Through harnessing the excellent capabilities of GPT{~\cite{brown2020language}} in interpreting the natural language content embedded within Swagger files for RESTful APIs, our proposed approach, {\tool} ({\tool}alon API Testing), leverages a powerful tool for comprehending intricate dependencies. This utilization of GPT empowers us to systematically extract and discern the dependencies that exist among the various API endpoints, indicating relationships that define the functionality of the APIs. Moreover, GPT possesses the capability to analyze the connections among the parameters associated with each operation. This enables us to perceive the underlying details of the API, ultimately leading to more thorough testing. In contrast, state-of-the-art methods, which mainly rely on heuristic approaches, fall short in capturing those dependencies. 

Our methodology employs a generative language model and advanced prompting techniques at key stages of the testing process. This includes (1) identifying relationships among schemas and operations, which are used to construct an operation dependency graph (ODG), and (2) detecting dependencies among input parameters as well as inter-dependencies between operations and their respective parameters for both valid and invalid test data generation. To streamline this process, our approach seamlessly integrates the ODG into the test script generation. The resulting test scripts are then combined with the generated test data to create a comprehensive suite of test cases, which are subsequently executed on the target APIs.


We have conducted an experiment to evaluate our approach, {\tool}, using a dataset comprising 12 RESTful API services. The results demonstrate a significant improvement of 15.7\% in the coverage of status codes documented in  OAS files over the state-of-the-art RestTestGen~\cite{corradini2022resttestgen}. Additionally, {\tool} effectively identifies the status codes that are not explicitly specified in the OAS files while reducing the number of false-positive test cases generated.

In this paper, we make the following key contributions:

1) {\tool}, an GPT-based approach to generate tests for RESTful API testing that considers the {\em inter-dependencies among operations, inter-parameter dependencies, and the dependencies between operations and parameters}.

2) An extensive evaluation showing our approach outperforming the state-of-the-art RESTful API testing approaches.
\section{Motivating Example}
\label{sec:motiv}

\subsection{Observations}

In this section, we use an example to elucidate the issue and inspire our approach. Fig.~\ref{fig:example} shows a representative OAS file delineating the RESTful APIs for an online flight booking system. Within this OAS file, two endpoints are documented: \code{GET /flights} and \code{POST /booking}. The former endpoint serves the purpose of retrieving a list of available flights within a specified date range, while the latter facilitates passengers in reserving a new flight. For each endpoint (operation), the file contains a descriptive summary (e.g., lines 7 and 20), the list of parameters (lines 21--24) and the request body object (line 25) with their names and textual descriptions in the schema (e.g., lines 25--41). Moreover, there exist supplementary endpoints supporting pertinent operations (omitted here for brevity).

\begin{figure}[t]
\centering
\lstset{
    basicstyle=\ttfamily\scriptsize\color{blue},  
        frame=single,         
        rulecolor=\color{black},
        numbers=left,       
        captionpos=b,       
        morekeywords={get, post, put, delete, responses, parameters, requestBody, description, name, summary, content, schema, openapi, info, paths},
        keywordstyle=\color{blue},
        comment=[l]{:},
        commentstyle=\color{black},
        stringstyle=\color{},
        tabsize=1,
        xleftmargin=1.5em,
        framexleftmargin=1.5em,
        numbersep= 5pt,
        numberstyle=\color{black}\tiny,
        breaklines=true,    
        breakatwhitespace=false, 
        lineskip={-1pt},
        backgroundcolor=\color{mygray},
    }
\begin{lstlisting}[]
openapi: 3.0.0
info:
  title: Flight Booking API
paths:
  /flights:
    get:
      summary: Get Flights
      responses:
        '200':
          description: A list of available flights.
          content:
            application/json:
              schema:
                type: array
                items:
                  $ref: '#/components/schemas/Flight'

  /booking:
    post:
      summary: Book a new Flight
      parameters:
        - name: flightId
          schema:
            type: integer
      requestBody:
        content:
          application/json:
            schema:
              properties:
                departureDate:
                  type: string
                  format: date
                  description: format in YYYY-MM-DD. Should be after today.
                arrivalDate:
                  type: string
                  format: date
                  description: format in YYYY-MM-DD. Should be after `departureDate`.
                passengerName:
                  type: string
                passengerAge:
                  type: integer
      responses:
        '200':
          description: The booking is successful.
          content:
            application/json:
              schema:
                $ref: '#/components/schemas/Booking'

components:
  schemas:
    Flight:
      type: object
      properties:
        id:
          type: integer
        origin:
          type: string
        destination:
          type: string
    Booking:
      type: object
      properties:
        flight:
          $ref: '#/components/schemas/Flight'
        departureDate:
          type: string
          format: date
        arrivalDate:
          type: string
          format: date
        passengerName:
          type: string
        passengerAge:
          type: integer
\end{lstlisting}
\vspace*{-3mm}
\caption{An Example of OpenAPI/Swagger Specification (OAS) file}
\vspace*{-5mm}
\label{fig:example}
\end{figure}

From this example, we make the following observations:

\vspace{2pt}
\begin{Observation}[{\bf Dependencies among Operations}] 
{\em The usage of operations involves the dependencies among them.}
\label{obs:odg}
\vspace{-4mm}
\end{Observation}

A test case of an endpoint might only be able to be successfully tested if the prerequisite test cases of other endpoints have already been properly executed beforehand. This is because responses to these prerequisite endpoints may affect the request for a current endpoint under test. 
    
In the provided OAS (refer to Fig.~\ref{fig:example}), it is crucial to supply a proper parameter for accurate testing of the \code{POST /booking} endpoint. This parameter should consist of a single field named \code{flightId}, representing an existing flight in the database. To meet this condition, it is necessary to first successfully execute the \code{GET /flights} endpoint. This action will result in a response containing a list of available flights. From this list, a specific flight element can be chosen, and its corresponding \code{flightId} value extracted. This value is then integrated into the test case for the \code{POST /booking} operation.

\vspace{2pt}
\begin{Observation}[{\bf Dependencies among Parameters}] 
{\em The usage of an operation in a library involves the dependencies among its parameters.}
\label{obs:interparam}
\end{Observation}
    
Dependencies also exist among the parameters of an operation. Inter-parameter dependency refers to a constraint that exists between two or more input parameters of an API endpoint. This constraint may pertain to the required values of these parameters, which must satisfy a predefined condition. Alternatively, it could involve the presence of an optional parameter, rendering the absence of another parameter problematic. For instance, consider the endpoint \code{POST /booking}. Here, a tangible inter-parameter constraint is evident between \code{departureDate} (line 30) and \code{arrivalDate} (line 34), an association discernible to users. Naturally, it is expected that the departure date of a flight precedes its arrival date. Thus, when handling the result, the requesting parameters should account for this constraint to ensure a valid response.

\vspace{2pt}
\begin{Observation}[{\bf Dependencies between An Operation and Its Parameters}] 
{\em The usage of an operation involves the dependencies between that operation and its parameters.}
\label{obs:singleparam}
\end{Observation}
    
Certain endpoints may encompass fields with values that must adhere to constraints, requiring those values to make sense in a real-world context. For instance, within the OAS file in Fig.~\ref{fig:example}, the value of \code{passengerAge} specified in the \code{POST /booking} endpoint (line 40) must exceed zero, although this constraint is not explicitly expounded within the OAS file. Furthermore, the value assigned to \code{departureDate} must refer to a future date. Unlike the \code{passengerAge} parameter, this constraint is articulated in natural-language description.

\subsection{The state-of-the-art Approaches}
Despite that the dependencies exist among the endpoints and parameters as observed, the state-of-the-art API testing approaches do not sufficiently address and capture them.

Regarding the inter-operation dependencies, RestTestGen employs {\em a heuristic approach based on name matching}. This determination hinges on the presence of a shared field between the output of one operation (i.e., endpoint) and the input of another. For instance, in the OAS file in Fig.~\ref{fig:example}, the \code{GET /flights} endpoint yields a set of Flight objects, each featuring an \code{id} field. Conversely, the \code{POST /booking} endpoint anticipates a \code{flightId} field as part of its input. This disparity in field nomenclature implies that the heuristic algorithm might encounter difficulty in establishing dependencies or might potentially mis-attribute them to other endpoints sharing similar field names. Relying solely on field names, it may not correctly identify the \code{GET /flights} endpoint as a dependent operation of the \code{POST /booking} endpoint.

When it comes to inter-parameter dependencies, previous approaches have indeed taken them into account during the test case generation. However, it is important to acknowledge certain limitations in these methods. For example, RESTest~\cite{MartinLopez2021Restest} addresses this issue by requiring {\em manual inclusion of dependencies among parameters} in the testing OAS file, utilizing the \code{x-dependencies} field. This manual intervention demands a considerable amount of time from testers. 
Regarding the dependencies between the operations and parameters, the state-of-the-art approaches use heuristic-based (RestTestGen) or rule-based (bBOXRT) methods to generate valid and invalid test data for testing, but they may not fully address these constraints described in natural language or require manual intervention to specify the dependencies.

\subsection{Capability of GPT and Motivation}

The Large Language Models (LLMs) have been achieving remarkable successes in understanding texts and source code~\cite{brown2020language}.
In this work, to capture such dependencies, we propose to explore that capability of the LLMs. To motivate this point, we made a few prompts to GPT-3.5~\cite{brown2020language} to explore its capability in understanding the intricate dependencies among the API endpoints (or operations) and their parameters. In particular, we use the version of \code{gpt-3.5-turbo-1106} via the provided API from OpenAI to request. We first loaded the Swagger file in Fig.~\ref{fig:example} and constructed prompts. The following conversations with GPT show how it can address the dependency~problem:

\begin{tcolorbox}[colback=black!5!white, colframe=black, boxrule=1pt, left=0.5em, right=0.5em, top=0.5em, bottom=0.5em]
{\footnotesize
    {\bf Prompt}:
    {\em ``Are there any dependencies between the POST /booking operation and GET /flights operation?''}
    
    {\bf GPT's response}:
    {\em ``Yes, there is a dependency between the POST /booking operation and the GET /flights operation. Before making a booking, the user needs to retrieve the available flights by calling the GET /flights operation. This is necessary to obtain flightId, which is required in the request body of the POST /booking operation. Thus, the GET /flights operation is a prerequisite operation that is called before the POST /booking operation can be tested successfully.''
    
    }
}
\end{tcolorbox}

    GPT is also able to detect inter-parameter dependencies:
\begin{tcolorbox}[colback=black!5!white, colframe=black, boxrule=1pt, left=0.5em, right=0.5em, top=0.5em, bottom=0.5em]
{\footnotesize
    {\bf Prompt}: 
    {\em ``Are there any inter-parameter dependencies mentioned in the POST /booking operation spec?''}
    
    {\bf GPT's response}:
    {\em ``Yes, there are inter-parameter dependencies mentioned in the POST /booking operation spec. The arrivalDate parameter is dependent on the departureDate parameter. The description states that the arrivalDate should be after the departureDate. Therefore, the departureDate needs to be provided before the arrivalDate in order to ensure that the request is valid."
    
    }
}
\end{tcolorbox}

Finally, GPT is capable of understanding the constraints among the parameters of the operations:

\begin{tcolorbox}[colback=black!5!white, colframe=black, boxrule=1pt, left=0.5em, right=0.5em, top=0.5em, bottom=0.5em]
{\footnotesize
    {\bf Prompt}: 
    {\em ``Are there any constraints of the value of parameters appeared in the specification, which are not defined in the specification but worth noting in a real-world context?''}
    
    {\bf GPT's response}: 
    {\em ``In the provided specification, there are no specific constraints for the parameter values. However, in a real-world context, the following constraints are worth noting:
    
    1. departureDate: Must be a future date (YYYY-MM-DD) and cannot be in the past.
    
    2. arrivalDate: Must be after the departureDate (YYYY-MM-DD) and cannot be the same as or before the departureDate.
    
    3. passengerAge: Subject to minimum and maximum age requirements, age categories for fare differentiation, or legal age restrictions for specific flights or destinations.
    
    These constraints ensure the validity of flight and passenger details, aligning with the considerations for flight bookings."
    
    }
}
\end{tcolorbox}

In brief, GPT is capable of identifying the relationships between operations and their parameters on a token level. Thus, we leverage GPT in producing the test cases that cover those dependencies among the APIs and/or their parameters.
\section{Dependency-aware Automated API Testing}

\subsection{Approach Overview}
\label{subsec:overview}

\begin{figure}[htbp]
  \begin{minipage}{\columnwidth}
      \centering     
      \includegraphics[width=3.2 in]{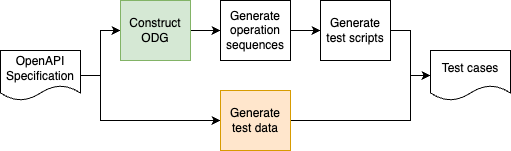}
      \caption{{\tool}: Dependency-aware Automated API Testing}
      \label{fig:approach overview}
  \end{minipage}
\end{figure}

This section presents an overview of our approach, {\tool} (Fig.~\ref{fig:approach overview}) that leverages LLMs, GPT in particular, to construct ODGs and generate test scripts and data for API testing. {\tool} receives the OAS file as input. This file serves as the basis for extracting the detailed information about its service to construct ODG and generate test scripts and data.
{\tool} consists of the following steps:

\begin{itemize}

    \item \textit{Construct ODG}: this step aims to construct an ODG representing dependencies between operations defined in the specification. See more details  in Section~\ref{subsec:odg}.
    \item \textit{Generate operation sequences}: using the ODG obtained from the previous step, our approach generates sequences of operation executions or requests in which one request prepares the data (parameters and request body) needed for succeeding requests. Dependent on specific ODG, it is possible that an operation is standalone or independent, which does not require any preceding requests to prepare the data needed for its successful execution.
    \item \textit{Generate test scripts}: this step aims to generate test scripts using the obtained operation sequences.
    \item \textit{Generate test data}: details of this step, which generates the data for a test operation, are given in Section~\ref{subsec:test data generation}.
\end{itemize}

\begin{figure}[t]
  \begin{minipage}{\columnwidth}
    \centering
    
    \includegraphics[width=0.8\textwidth]{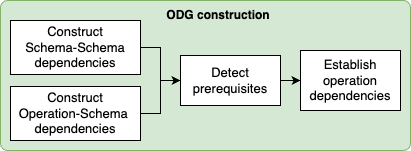}
    \caption{ODG Construction}
    \label{fig:ODG construction}
  \end{minipage}
\end{figure}

\subsection{Operation Dependency Graph Construction (Fig.~\ref{fig:ODG construction})}
\label{subsec:odg}


\subsubsection{Important Concepts}

\vspace{2pt}
\begin{Definition}[{\bf Operation Dependency Graph}] {\em An ODG is represented as a directed graph $G=(N,V)$, wherein each node $N$ signifies a discrete operation within the RESTful API. The presence of an edge $v \in V$, articulated as $v = n_1 \rightarrow n_2$, signifies the existence of a dependency between the nodes $n_1$ and $n_2$. This dependency is predicated on the condition that one or more fields in the output of source node $n_1$ must coincide with in the input of target node $n_2$, thereby mandating that the execution of operation $n_1$ precedes that of~$n_2$.} 
\end{Definition}

\begin{figure}[htpt]
  \begin{minipage}{\columnwidth}
      \centering
      \includegraphics[width=0.5\textwidth]{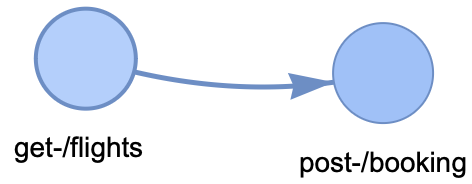}
      \vspace{-3pt}
        \caption{Dependence between \code{GET/flights} and \code{POST/booking} in ODG}
      \label{fig:odg ex}
  \end{minipage}
\end{figure}

\vspace{2pt}
The concept of ODG was previously explored in the work of RestTestGen. Refer to the exemplar OAS file in Fig.~\ref{fig:example}, and observe a dependency between the two endpoints, \code{GET /flights} and \code{POST /booking}, as explained in Observation~\ref{obs:odg} (Fig.~\ref{obs:odg}). Part of the ODG of this OAS file is shown in Fig.~\ref{fig:odg ex}. 

\subsubsection{ODG Construction Algorithm}
\label{subsec:odg:construct algo}

Our algorithm differs from the heuristic-based ODG construction algorithm in RestTestGen~\cite{viglianisi2020resttestgen} in its utilization of GPT. Algorithm~\ref{algo:odg} presents our pseudocode, which involves the use of GPT for inferring the dependencies between operations and schemas, as well as the dependencies among schemas. These process steps are employed to construct additional operation dependency edges that heuristic-based algorithms cannot detect.

\newcommand\mycommfont[1]{\scriptsize\ttfamily\textcolor{blue}{#1}}
\SetCommentSty{mycommfont}

\begin{algorithm}[h]
\footnotesize

\SetAlgoLined
\SetKwFunction{generateOperationDependencies}{generateOperationDependencies}
\SetKwFunction{GPTgenSchemaSchemaDep}{GPTgenSchemaSchemaDep}
\SetKwFunction{GPTgenOperationSchemaDep}{GPTgenOperationSchemaDep}
\SetKwFunction{getAllOperationsFrom}{getAllOperationsFrom}
\SetKwFunction{getParameters}{getParameters}
\SetKwFunction{getRelevantSchemas}{getRelevantSchemas}
\SetKwFunction{getDep}{getDep}
\SetKwFunction{findPrecedingOperations}{findPrecedingOperations}
\SetKwFunction{extendDependencies}{extendDependencies}
\SetKwFunction{gatherOperationDep}{gatherOperationDep}
\SetKwFunction{gatherODheuristic}{gatherODheuristic}

\SetKwProg{Function}{Function}{}{}
\SetKwInOut{Output}{Output}
\SetKwInOut{Input}{Input}

\Function{\generateOperationDependencies{\textrm{Swagger}}}{
    \Input{A Swagger Specification file}
    \Output{List of Operation Dependencies $\textit{\textbf{OD}}$} 
    $\textit{\textbf{OD}}$ $\leftarrow$ \gatherODheuristic(Swagger) \label{algo:odg:manual}\\
    $\textit{\textbf{OS}}$ $\leftarrow$ \GPTgenOperationSchemaDep(Swagger) \label{algo:odg:osdeps} \\
    $\textit{\textbf{SS}}$ $\leftarrow$ \GPTgenSchemaSchemaDep(Swagger) \label{algo:odg:ssdeps}\\
    $\textit{\textbf{O}}$ $\leftarrow$ \getAllOperationsFrom(Swagger) \label{algo:odg:getops}\\
     \ForEach{$\textbf{O}_i$ \textbf{in} \textbf{O}\label{algo:odg:for1}} {
        $\textit{\textbf{P}}$ $\leftarrow$ \getParameters($\textit{\textbf{O}}_i$) \label{algo:odg:getparam} \\
        $\textit{\textbf{PSDep}}$ $\leftarrow$ \getDep($\textit{\textbf{OS}}$, $\textit{\textbf{O}}_i$) \label{algo:odg:getdep} \\
        \If{$\textit{\textbf{PSDep}}$ $\neq$ $\varnothing$}{
            $\textit{\textbf{NewOD}}$ $\leftarrow$ \gatherOperationDep($\textit{\textbf{PSDep}}$,$\textit{\textbf{P}}$) \label{algo:odg:gather1}\\
            $\textit{\textbf{OD}}$.extend($\textit{\textbf{NewOD}}$) \label{algo:odg:extendod}
        }
        \If{$\textit{\textbf{P}}$ $\neq$ $\varnothing$}{ \label{algo:odg:param not none}
            $\textit{\textbf{ChildS}}$ $\leftarrow$ \getDep($\textit{\textbf{SS}}$, $\textit{\textbf{O}}_i$) \label{algo:odg:getdep2}\\
            $\textit{\textbf{NewOD}}$ $\leftarrow$ \gatherOperationDep($\textit{\textbf{ChildS}}$,$\textit{\textbf{P}}$)\label{algo:odg:gather2}\\
            $\textit{\textbf{OD}}$.extend($\textit{\textbf{NewOD}}$) \label{algo:odg:extend2}
        }   
    }
    \label{algo:odg:endfor1}
    \KwRet $\textit{\textbf{OD}}$
}
\Function{\gatherOperationDep($\textit{\textbf{C}}, \textit{\textbf{P}}$\label{algo:odg:fgather})}{
    \Input{A collection to find new OD elements $\textit{\textbf{C}}$\\
    A set of parameters to check dependency $\textit{\textbf{P}}$}
    \Output{A set of new elements to be extended $\textit{\textbf{NewOD}}$}
    \ForEach{$\textit{\textbf{C}}_j$ \textbf{in} $\textit{\textbf{C}}$\label{algo:odg:for2}}{
        $\textit{\textbf{PO}}$ $\leftarrow$ \findPrecedingOperations($\textit{\textbf{C}}_j$) \label{algo:odg:precede}\\
        \If{$\textit{\textbf{PO}}$ $\neq$ $\varnothing$ $\vee$ $\textit{\textbf{PO}} \in \textit{\textbf{P}}$} {
       \extendDependencies($\textit{\textbf{NewOD}}$, $\textit{\textbf{PO}}$) \label{algo:odg:extenddep}
        }
$\textit{\textbf{P}}$.remove($\textit{\textbf{PO}}$)
    }   
    \KwRet $\textit{\textbf{NewOD}}$
}

\caption{ODG Construction}
\label{algo:odg}
\end{algorithm}

\vspace{2pt}
\subsubsection*{\bf \em Heuristic-based collection of edges}

Primarily, the function \code{gatherODheuristic} (line~\ref{algo:odg:manual}) is employed to iterate through all operations, attempting to match input-output pairs. Upon detecting a pair with a perfect character match, we add the resulting edge between the corresponding nodes in the graph.

In the case of the OAS file depicted in Fig.~\ref{fig:example}, the relationship between two operations is considered. Although the pair \code{flightId} (line 22, input of \code{POST /booking}) and \code{id} (existed in the schema ``Flight", the output of \code{GET /flights}, on line 16) do not exhibit a perfect match, a significant dependency persists due to the presence of \code{id} within the ``Flight" schema, semantically aligning with \code{flightId}. To resolve this, we use GPT to analyze such pairs in a two-step process, involving the establishment of Operation-Schema dependencies (line~\ref{algo:odg:osdeps}) and Schema-Schema dependencies (line~\ref{algo:odg:ssdeps}), thereby acting as a bridge to identify the dependencies between operations.

\vspace{3pt}
\subsubsection*{\bf \em Operation-Schema Dependency}

\vspace{1pt}
\begin{Definition}[{\bf Operation-Schema Dependency}]
{\em The origin of two operations depends on each other when the response of one operation is required as input for the successful execution of the other. Within an OAS, each operation's response is defined under a schema, and subsequent operations must identify relevant schemas to determine their predecessor operations. An operation-schema dependency encompasses all schemas that share keys between their fields and the input parameters of the operation, which GPT can effectively identify.}  
\end{Definition}

\vspace{1pt}
Operation-Schema dependencies capture this relationship by constructing the dictionary \textit{\textbf{OS}} (line 3). In this dictionary, each key corresponds to an operation’s name $o_i \in O_n$, in~which $O_n$ is a set of operation’s name described in the~OAS specification. This key is associated with a collection of data, providing insights into essential prerequisite schemas and the connections between the current operation $o_i$ and these schemas through their parameters. For each parameter pair, the first property belongs to the schema, while the second property belongs to the operation $o_i$. As an example, consider the $\textit{\textbf{OS}}$ dictionary created for the OAS file shown in Fig.~\ref{fig:example}, which is illustrated in Fig.~\ref{fig:os ex}. Fig.~\ref{fig:gpt convers os} shows the prompt we used to infer all prerequisite schemas of the operation \code{POST /booking}.

\begin{figure}[t]
\colorlet{punct}{red!60!black}
\definecolor{background}{HTML}{EEEEEE}
\definecolor{delim}{RGB}{20,105,176}
\colorlet{numb}{magenta!60!black}
\lstset{
    basicstyle=\scriptsize\ttfamily,
    frame=lines,   
    rulecolor=\color{black},
    numbers=left,       
    stepnumber=1,
    numbersep=8pt,
    captionpos=b,    
    tabsize=2,
    xleftmargin=1.5em,
    framexleftmargin=1.5em,
    numbersep= 5pt,
    numberstyle=\color{black}\tiny,
    breaklines=true,    
    breakatwhitespace=false, 
    lineskip={-1pt},
    showstringspaces=false,
    literate=
     *{0}{{{\color{numb}0}}}{1}
      {1}{{{\color{numb}1}}}{1}
      {2}{{{\color{numb}2}}}{1}
      {3}{{{\color{numb}3}}}{1}
      {4}{{{\color{numb}4}}}{1}
      {5}{{{\color{numb}5}}}{1}
      {6}{{{\color{numb}6}}}{1}
      {7}{{{\color{numb}7}}}{1}
      {8}{{{\color{numb}8}}}{1}
      {9}{{{\color{numb}9}}}{1}
      {:}{{{\color{punct}{:}}}}{1}
      {,}{{{\color{punct}{,}}}}{1}
      {\{}{{{\color{delim}{\{}}}}{1}
      {\}}{{{\color{delim}{\}}}}}{1}
      {[}{{{\color{delim}{[}}}}{1}
      {]}{{{\color{delim}{]}}}}{1},
    }
\begin{lstlisting}
OS = {
  "post-/booking": {
    "Flight": {
      "flightId": "id"
    }
    "Booking": {
      "flightId": "flight"
    }
  }
}
\end{lstlisting}
\vspace*{-3mm}
\caption{Example of an Operation-Schema dependency dictionary. The operation \code{POST /booking} has two Operation-Schema dependencies with the schemas ``Flight" and ``Booking". This is indicated by the pairs of parameter \code{flightId} (line 22) with field \code{id} (line 55) and \code{flight} (line 64) in Fig~\ref{fig:example}.}
\label{fig:os ex}
\end{figure}

\begin{figure}[t]
\definecolor{mycolor}{RGB}{20,105,176}
\centering
\lstset{
    basicstyle=\ttfamily\scriptsize,  
        frame=lines,         
        rulecolor=\color{black},
        numbers=left,       
        captionpos=b,       
        commentstyle = \color{mycolor}, 
        keywordstyle = \color{blue}, 
        stringstyle = \color{red}, 
        tabsize=2,
        xleftmargin=1.5em,
        framexleftmargin=1.5em,
        numbersep= 5pt,
        numberstyle=\color{black}\tiny,
        breaklines=true,    
        breakatwhitespace=false, 
        lineskip={-1pt}
    }
\begin{lstlisting}[language=python]
PROMPT="""Given the operation and its parameters, identify all prerequisite 
schemas for retrieving information related to the operation's parameters.

Below is the operation and its parameters:
post-/booking:
  flightId: integer
  departureDate: string
  arrivalDate: string
  ...

Below is the list of all schemas and their properties:
schemas:
  Flight:
    ...
  Booking:
    ...

Please format the prerequisite schemas in the following structure:
<parameter of the operation> -> <equivalent operation of the relevant schema>
..."""

\end{lstlisting}
\vspace*{-3mm}
\caption{The prompt for prerequisites in the \code{POST /booking} operation yields a GPT response with the ``Flight" and ``Booking" schemas, creating Operation-Schema dependencies.}
\label{fig:gpt convers os}
\end{figure}

Fig.~\ref{fig:os ex} shows the operation \code{POST /booking} along with two schemas, ``Flight'' and ``Booking,'' identified as its prerequisite schemas by GPT. When detecting the ``Flight'' schema, it allows to infer that \code{GET /flights} is the~predecessor since the ``Flight'' schema is specified as the reply for \code{GET /flights}. However, in some cases, GPT only identifies the ``Booking'' schema, and cannot identify a clear predecessor operation since there is no operation specifying ``Booking'' as its~response. To address this, we use a Schema-Schema~dependency.

\vspace{2pt}
\subsubsection*{\bf \em Schema-Schema Dependency}

\vspace{3pt}
\begin{Definition}[{\bf Schema-Schema Dependency}]
{\em We define $S_n$ is a set of $n$ schemas described in the OAS specification. Schema-schema dependencies model the relationship between schemas by initializing the dictionary \textit{\textbf{SS}} (line 4). This dictionary contains $n$ distinct key-value pairs where $key$ is a schema $s_i \in S_n$, and $value$ is a list of schemas, where:}
\begin{itemize}
    \item {\em Each schema in the list contains at least one field that refers to a field specified in the $key$, or}
    \item {\em The schema itself is referred by the $key$ (e.g. the element ``Flight" in the $value$ is referred by the $key$~``Booking").}
\end{itemize}
\end{Definition}

Consider the conversation displayed in Fig.~\ref{fig:gpt convers ss} with GPT. Our approach tries to obtain potential relationships of the schema ``Booking'' using the Swagger displayed in Fig.~\ref{fig:example}.

\begin{figure}[t]
\definecolor{mycolor}{RGB}{20,105,176}
\centering
\lstset{
    basicstyle=\ttfamily\scriptsize,  
    frame=lines,
    rulecolor=\color{black},
    numbers=left,
    captionpos=b,
    commentstyle=\color{mycolor},
    keywordstyle=\color{blue},
    stringstyle=\color{red},
    tabsize=2,
    xleftmargin=1.5em,
    framexleftmargin=1.5em,
    numbersep=5pt,
    numberstyle=\color{black}\tiny,
    breaklines=true,
    breakatwhitespace=false,
    lineskip={-1pt}
}
\begin{lstlisting}[language=python]
PROMPT="""Given the schema and its properties in the OpenAPI specification of an API application, your task is to identify the prerequisite schemas that need to be created before establishing the mentioned schema.

Below is the schema and its properties
Booking:
  flight: 
    $ref: '#/components/schemas/Flight'
  departureDate: ...
  arrivalDate: ...
  ...
 
Below is the list of all schemas and their properties:
schemas:
  Flight:
    ...
  Booking:
    ...

Return in separated lines. No explanation needed."""

\end{lstlisting}
\vspace*{-3mm}
\caption{The prompt for ``Booking" prerequisites elicits a GPT response with the ``Flight" schema, establishing a Schema-Schema dependency.}
\label{fig:gpt convers ss}
\end{figure}

After iterating through all schemas, the $\textit{\textbf{SS}}$ for the OAS file given in Fig.~\ref{fig:example} will have the value as shown below.

\colorlet{punct}{red!60!black}
\definecolor{background}{HTML}{EEEEEE}
\definecolor{delim}{RGB}{20,105,176}
\colorlet{numb}{magenta!60!black}
\lstset{
    basicstyle=\scriptsize\ttfamily,
    frame=lines,   
    rulecolor=\color{black},
    numbers=left,       
    stepnumber=1,
    numbersep=8pt,
    captionpos=b,    
    tabsize=2,
    xleftmargin=1.5em,
    framexleftmargin=1.5em,
    numbersep= 5pt,
    numberstyle=\color{black}\tiny,
    breaklines=true,    
    breakatwhitespace=false, 
    lineskip={-1pt},
    showstringspaces=false,
    literate=
     *{0}{{{\color{numb}0}}}{1}
      {1}{{{\color{numb}1}}}{1}
      {2}{{{\color{numb}2}}}{1}
      {3}{{{\color{numb}3}}}{1}
      {4}{{{\color{numb}4}}}{1}
      {5}{{{\color{numb}5}}}{1}
      {6}{{{\color{numb}6}}}{1}
      {7}{{{\color{numb}7}}}{1}
      {8}{{{\color{numb}8}}}{1}
      {9}{{{\color{numb}9}}}{1}
      {:}{{{\color{punct}{:}}}}{1}
      {,}{{{\color{punct}{,}}}}{1}
      {\{}{{{\color{delim}{\{}}}}{1}
      {\}}{{{\color{delim}{\}}}}}{1}
      {[}{{{\color{delim}{[}}}}{1}
      {]}{{{\color{delim}{]}}}}{1},
    }
\begin{lstlisting}
SS = {
  "Flight": [],
  "Booking": [
    "Flight"
  ]
}
\end{lstlisting}

\vspace{2pt}
\subsubsection*{\bf \em The process to detect prerequisites and construct edges of the graph}

Upon the completion of constructing these two hierarchical sets generated by the GPT model, we initiate a process to establish dependencies between operations.

The process has a \code{for} loop (lines \ref{algo:odg:for1}--\ref{algo:odg:endfor1}), iterating through extracted operations from the OAS file (line~\ref{algo:odg:getops}). The objective is to identify the relevant schema for each parameter obtained from the current operation (line~\ref{algo:odg:getparam}). Using the Operation-Schema dependency stored in $\textit{\textbf{OS}}$ (line~\ref{algo:odg:getdep}), $\textit{\textbf{OD}}$ is updated (line~\ref{algo:odg:extendod}). Any unaccounted parameters at the current schema level prompt an attempt to retrieve the next schema level using $\textit{\textbf{SS}}$ (line~\ref{algo:odg:getdep2}), followed by a second attempt to update the final result (line~\ref{algo:odg:extend2}). These steps establish the adequacy of $\textit{\textbf{OD}}$ to generate the sequential order of operations for test generation.

Specifically, the function \code{gatherOperationDep} (line~\ref{algo:odg:fgather}) takes two inputs. It contains a single loop (line~\ref{algo:odg:for2}), iterating through all elements in the collection. Each element is examined by \code{findPrecedingOperations} (line~\ref{algo:odg:precede}). This function returns a list of preceding operations (\textbf{\textit{PO}}), representing POST or GET operations, indicating the need for creating a new data item before executing it on any test cases or ensuring the existence of that data in the database. After this, $\textit{\textbf{P}}$ checks the edges to determine if $\textit{\textbf{PO}}$ is usable for the current operation.

\subsection{Test Data Generation}
\label{subsec:test data generation}

\begin{figure}[h]
  \begin{minipage}{\columnwidth}
      \centering \includegraphics[width=\textwidth]{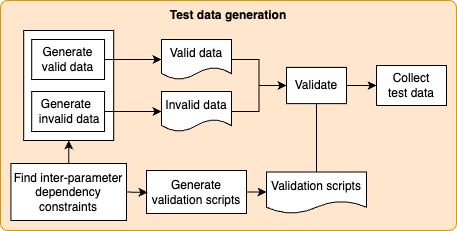}
      \caption{The process of test data generation for a single operation}
      \label{fig:test data generation}
  \end{minipage}
\end{figure}

This section describes our approach to leverage GPT to generate both valid and invalid test data, as well as Python scripts to validate the accurate correspondence between parameters.
We use GPT to generate high-quality valid and invalid test data that closely mimic real-world data. Additionally, GPT is used to identify inter-dependencies among the operation's parameters. These dependencies serve as context for dynamically update the data generation, and are used for formulating Python validation scripts. When creating test data files for failure cases, we consider various scenarios, such as omitting values in required fields, providing incorrect data types for certain fields, or violating inter-parameter constraints.


These Python validation scripts ensure consistency with the data file. However, if no inter-parameter dependencies are identified in the current operation, the Python script generation process does not take place, and no additional update in data generation. Following another cycle of data generation, each item in the valid/invalid data files undergoes an additional evaluation phase before being compiled into usable test data files. These files are then integrated with a test script to form a test case for the operation. JSON format is chosen for communication between a data item and its script.



All prompts follow the format shown in Fig.~\ref{fig:gpt data gen}. The section labeled \code{additional\_instruction} is employed to guide the model in generating valid or invalid datasets and to specify failed scenarios for more comprehensive coverage of failures.

\begin{figure}[t]
\definecolor{mycolor}{RGB}{20,105,176}
\centering
\lstset{
    basicstyle=\ttfamily\scriptsize,  
        frame=lines,         
        rulecolor=\color{black},
        numbers=left,       
        captionpos=b,       
        commentstyle = \color{mycolor}, 
        keywordstyle = \color{blue}, 
        stringstyle = \color{red}, 
        tabsize=2,
        xleftmargin=1.5em,
        framexleftmargin=1.5em,
        numbersep= 5pt,
        numberstyle=\color{black}\tiny,
        breaklines=true,    
        breakatwhitespace=false, 
        lineskip={-1pt}
    }
\begin{lstlisting}[language=python]
GET_DATASET_PROMPT=f"""Given the information about the operation, generate a 
dataset containing 10 data items to be used to test the operation.
{additional_instruction}

Operation information: {endpoint_information}
Referenced schema: {ref_schema}

Your dataset represents each data item in the JSONL format, line by line."""
\end{lstlisting}
\vspace*{-3mm}
\caption{Sample prompt template for test data generation}
\label{fig:gpt data gen}
\end{figure}

An illustration for the generated data, which represents the value of the request body for \code{POST /booking} is displayed in Fig. \ref{fig:example}. The valid and invalid data is as shown in Fig. \ref{fig:gen data}.

\begin{figure}[t]
\centering
\colorlet{punct}{red!60!black}
\definecolor{background}{HTML}{EEEEEE}
\definecolor{delim}{RGB}{20,105,176}
\colorlet{numb}{magenta!60!black}
\lstset{
    basicstyle=\scriptsize\ttfamily,
    frame=lines,         
    rulecolor=\color{black},
    numbers=left,       
    stepnumber=1,
    numbersep=8pt,
    captionpos=b,       
    tabsize=2,
    xleftmargin=1.5em,
    framexleftmargin=1.5em,
    numbersep= 5pt,
    numberstyle=\color{black}\tiny,
    breaklines=true,    
    breakatwhitespace=false, 
    lineskip={-1pt},
    showstringspaces=false,
    literate=
     *{0}{{{\color{numb}0}}}{1}
      {1}{{{\color{numb}1}}}{1}
      {2}{{{\color{numb}2}}}{1}
      {3}{{{\color{numb}3}}}{1}
      {4}{{{\color{numb}4}}}{1}
      {5}{{{\color{numb}5}}}{1}
      {6}{{{\color{numb}6}}}{1}
      {7}{{{\color{numb}7}}}{1}
      {8}{{{\color{numb}8}}}{1}
      {9}{{{\color{numb}9}}}{1}
      {:}{{{\color{punct}{:}}}}{1}
      {,}{{{\color{punct}{,}}}}{1}
      {\{}{{{\color{delim}{\{}}}}{1}
      {\}}{{{\color{delim}{\}}}}}{1}
      {[}{{{\color{delim}{[}}}}{1}
      {]}{{{\color{delim}{]}}}}{1},
    }
\begin{lstlisting}[]
[
    { 
        "data": {
            "departureDate": "2022-12-01",
            "arrivalDate": "2022-12-02",
            "passengerName": "John Doe",
            "passengerAge": 30
        },
        "expected_code": 200 
    },
    { 
        "data": {
            "departureDate": "2022-11-15",
            "arrivalDate": "2022-11-16",
            "passengerName": "Jane Smith",
            "passengerAge": 25 
        },
        "expected_code": 200 
    }, 
    ...
]
\end{lstlisting}
\begin{lstlisting}[]
[
    { 
        "data": {
            "departureDate": "2022-03-10",
            "arrivalDate": "2022-03-09",
            "passengerName": "Michael Johnson",
            "passengerAge": 35 
        },
        "expected_code": 400 
    },
    { 
        "data": {
            "departureDate": null,
            "arrivalDate": "",
            "passengerName": "John Doe",
            "passengerAge": "25" 
        },
        "expected_code": 400 
    }, 
    ...
]
\end{lstlisting}
\vspace*{-3mm}
\caption{These items illustrate valid and invalid request body data for the \code{POST /booking} endpoint. }
\label{fig:gen data}
\end{figure}

\subsection{Generating Test Scripts and Test Cases}

\begin{figure}[h]
\centering
\lstset{
    basicstyle=\ttfamily\scriptsize,  
        frame=lines,         
        rulecolor=\color{black},
        numbers=left,       
        captionpos=b,       
        commentstyle = \color{green}, 
        keywordstyle = \color{blue}, 
        morekeywords={def, import, makeRequest, assertStatusCode},
        stringstyle = \color{red}, 
        tabsize=2,
        xleftmargin=1.5em,
        framexleftmargin=1.5em,
        numbersep= 5pt,
        numberstyle=\color{black}\tiny,
        breaklines=true,    
        breakatwhitespace=false, 
        lineskip={-1pt}
    }
\begin{lstlisting}[]
// Import statements
import ...

// ChatGPT generated test data goes here
def path_variables_1 = [:]
def query_parameters_1 = [:]
def body_1 = ""

def response_1 = makeRequest(
    <path to test GET /flight>, path_variables_1, 
    query_parameters_1, body_1
)

// ChatGPT generated test data goes here
def path_variables = [:]
def query_parameters = [
  'flightId': response_1[0].id
]
def body = <get from the relevant data file>

def response = makeRequest(
    <path to test POST /booking>, path_variables, 
    query_parameters, body
)
def latest_response = response

// END
def expected_status_code = <value>
assertStatusCode(latest_response, expected_status_code)
\end{lstlisting}
\vspace*{-3mm}
\caption{The Groovy-generated test script exemplifies the \code{POST /booking} operation following the \code{GET /flight} $\rightarrow$ \code{POST /booking} sequence.}
\label{fig:groovy}
\end{figure}

We adhere to the approach of examining response status codes within the 2xx and 4xx ranges. For an operation, a corresponding test script is generated by the GPT model, based on the number of sequences extrapolated from the ODG. The leading operation in each order retrieves the data element from its relevant data file, and its output is assimilated as the input of the successive operation in the sequential procession. Each operation is furnished with a tailored test script that exclusively draws a data element from the test data file, housing only a singular execution of that operation in and of itself.

Fig.~\ref{fig:groovy} displays a generated Groovy programming language test script for consecutively executing two requests to test the operation \code{POST /booking}, as described in Fig.~\ref{fig:example}. It illustrates how the operation dependencies impact the order of execution of different operations in a single script. Custom keywords like \code{makeRequest} and \code{assertStatusCode} have been created to facilitate clean Groovy code, making it easier for the GPT model to generate an executable test script. 

After establishing a collection for evaluating 2xx status codes, we used these scripts as templates to create scripts for assessing 4xx status codes. These existing scripts can be adjusted by modifying the path of relevant test data files and rearranging the order within the operation sequence. For example, if a pre-existing DELETE operation aligns with the current element in the sequence, it can be inserted to simulate a scenario of receiving a 404 status code.


\section{Empirical Evaluation}
\label{sec:eval}

\subsection{Research Questions}

For evaluation, we seek to answer the following questions:

\indent {\bfseries RQ1.} {\bfseries [Test coverage]}: How well does our approach generate valid test cases for coverage improvement? \\
\indent {\bfseries RQ2.} {\bfseries [Test generation efficiency]}: How efficient is our approach in generating valid test cases? \\
\indent {\bfseries RQ3.} {\bfseries[Failure detection]}: What is the capability of our approach in detecting failures and mismatches between the working API and its specification?

We chose to compare our approach against the state-of-the-art RestTestGen (RTG for short).


For RQ1, our analysis involves using RTG and our approach (KAT) to generate test cases and run them on the target APIs to measure the test coverage of successful (2xx) and operation failure responses (4xx). 


For RQ2, we evaluate the efficiency of the approaches by measuring the efficiency score which is the number of test cases triggering 2xx and 4xx responses over the total number of test cases generated. A higher efficiency score indicates a more efficient approach, which also means fewer test cases needed to generate for covering status codes. 

For RQ3, we measure the number of 500 errors found, the number of status codes not documented in the OAS, and the number of mismatches in status codes defined in the OAS and those returned while making requests to target services. 

\subsection{Dataset}
The dataset used in our experimentation consists of 12 RESTful API services. We decided to use these services for several reasons (1) they are publicly accessible to allow replication, including Canada Holidays and the Bills API collected from the website APIs.guru~\cite{apiguru}, (2) most of them are previously investigated in prior studies, including PetStore in \cite{viglianisi2020resttestgen, liu2022morest,corradini2021restats,vassiliou2020simple}, Genome Nexus in \cite{kim2022automated, kim2023adaptive,kim2023enhancing}, BingMap-Route in \cite{wu2022combinatorial,corradini2022resttestgen}, GitLab in \cite{yamamoto2021efficient,wu2022combinatorial, atlidakis2019restler,karlsson2020quickrest,lin2022forest} (3) ProShop~\cite{proshop}, an in-house development, is a private service with an OAS file that lies beyond the scope of GPT's knowledge, ensuring fair comparison (4) they represent a diverse set of services from different domains with various status codes, operation dependencies, and parameter constraints. 

Table \ref{tab:dataset_stat} provides a summary of the services used in our experiments. \textit{OP} is the number of operations defined in the OAS file for the service. \textit{2xx SC} and \textit{4xx SC} denote 
the counts of 2xx status codes and 4xx codes for all operations, respectively. \textit{Par} is the total number of parameters, including those within the request bodies of operations. \textit{DepOps} describes the number of operations with dependencies, and \textit{IntOps} specifies the number of operations with at least one inter-parameter constraint.

\begin{table}[t]
  \centering
    \caption{Statistics of REST service dataset}
    \label{tab:dataset_stat}
    \begin{tabular}{|P{2.2cm}|*{6}{P{0.55cm}|}}
     \hline
     REST service & OP & 2xx SC & 4xx SC & Par & Dep Ops & Int Ops \\
     \hline
     ProShop & 16 & 16 & 30 & 46 & 11 & 0\\
     \hline
     PetStore & 19 & 15 & 20 & 61 & 10 & 0 \\
     \hline
     Canada Holidays & 6 & 6 & 2 & 9 & 2 & 0 \\
     \hline
     Bills API & 21 & 21 & 31 & 57 & 13 & 0 \\
     \hline
     Genome Nexus & 23 & 23 & 0 & 44 & 10 & 0 \\
     \hline
     BingMap-Route & 14 & 14 & 0 & 162 & 1 & 10 \\
     \hline
     GitLab-Branch  & 9 & 9 & 0 & 88 & 7 & 1 \\
     \hline
     GitLab-Commit & 15 & 15 & 0 & 135 & 13 & 2 \\
     \hline
     GitLab-Groups & 17 & 17 & 0 & 152 & 15 & 0 \\
     \hline
     GitLab-Issues & 27 & 27 & 0 & 239 & 22 & 9 \\
     \hline
     GitLab-Project & 31 & 31 & 0 & 295 & 29 & 4 \\
     \hline
     GitLab-Repository & 10 & 10 & 0 & 101 & 8 & 2 \\
     \hline
     Average & 17.3 & 17 & 6.9 & 115.8 & 11.8 & 2.3 \\
     \hline
    \end{tabular}
\end{table}

\section{Performance on Test coverage (RQ1)}

\subsection{Experimental Methodology}

In the experiment, we applied both RTG (version 23.09) and our approach to all 12 services in the collected dataset (Table \ref{tab:dataset_stat}). For each service, we analyzed the results of test case execution, with a specific focus on the final response status code of each generated test case.

To address RQ1, we formulated as follows:

\indent \textbf{2xx Coverage}: It evaluates test coverage for successful responses by calculating the ratio of total 2xx response codes documented and triggered by at least one test case to the number of documented status codes within the 2xx range.

\indent \textbf{4xx Coverage}: This assesses test coverage of failure responses by calculating the ratio of total 4xx response codes documented and triggered by at least one test case to the number of documented status codes within the 4xx range.

\indent \textbf{Overall Coverage}: This measures the ratio of total actual response codes to the total documented status codes.

\indent \textbf{Average}: We determine the average coverage for each measure by calculating the ratio of the total actual response codes across all services to the total number of status codes documented in the OAS files.

\subsection{Experimental Results}

\begin{table}[t]
  \centering
    \caption{Documented status code coverage (RQ1)}
    \label{tab: testcoveragequality}
    \begin{tabular}{|P{2.2cm}|*{6}{P{0.55cm}|}}
     \hline
     \multicolumn{1}{|c|}{\multirow{2}{*}{REST service}} & \multicolumn{2}{c|}{\makecell{Overall \\ coverage (\%)}} & \multicolumn{2}{c|}{\makecell{2xx \\ coverage (\%)}} & \multicolumn{2}{c|}{\makecell{4xx \\ coverage (\%)}} \\
     \cline{2-7}
     & RTG & KAT & RTG & KAT & RTG & KAT \\
     \hline
     ProShop & 67.4 & {\bfseries 87} & 62.5 & {\bfseries 100} & 70 & {\bfseries 80} \\
     \hline
     Petstore & 54 & {\bfseries 74.3} & 66.7 & {\bfseries 100} & 50 & {\bfseries 55} \\
     \hline
      Canada Holidays & 100  & 100 & 100 & 100 & 100 & 100 \\
     \hline
     Bills API & 71.2 & {\bfseries 82.7} & 66.7 & {\bfseries 81} & 74.2 & {\bfseries 83.9} \\
     \hline
     Genome Nexus & 78.3 & {\bfseries 100} & 78.3 & {\bfseries 100} & - & - \\
     \hline
     BingMap-Route & 28.6 & {\bfseries 78.6} & 28.6 & {\bfseries 78.6} & - & - \\
     \hline
     GitLab-Branch & 55.6 & {\bfseries 77.8} & 55.6 &  {\bfseries 77.8} & - & - \\
     \hline
     GitLab-Commit & 20 & 20 & 20 & 20 & - & - \\
     \hline
     GitLab-Groups & 52.9 & {\bfseries 58.8} & 52.9 & {\bfseries 58.8} & - & - \\
     \hline
     GitLab-Issues & 40.7 & {\bfseries 63} & 40.7 & {\bfseries 63} & - & - \\
     \hline
     GitLab-Project & 67.7 & {\bfseries 74.2} & 67.7 & {\bfseries 74.2} & - & - \\
     \hline
     GitLab-Repository & 40 & 40 & 40 & 40 & - & - \\
     \hline
     Average & 59.2 & {\bfseries 74.9} & 56.4 & {\bfseries 74.5} & 67.5 & {\bfseries 75.9} \\
     \hline
     Standard deviation & 21.4 & 22.3 & 21.2 & 24.5 & 17.8 & 16.1 \\
     \hline
    \end{tabular}
\end{table}

Table \ref{tab: testcoveragequality} displays the coverage result. Our observations show an enhancement in the average coverage metrics over the state-of-the-art RTG. Specifically, we noted a 15.7\% increase in overall coverage, a 18.1\% improvement in 2xx coverage, and a 8.4\% increase in 4xx coverage, on average, over RTG.

We observed that for the BingMap Route service, which has the highest number of operations containing inter-parameter dependency constraints, {\tool} exhibits a high improvement of up to 50\% in both overall coverage and 2xx coverage. Unlike RTG, which assigned values without considering inter-parameter dependency constraints, {\tool}, powered by GPT, effectively identifies and accommodates these constraints, generating valid test data that align with the detected constraints. In GitLab subsystems, particularly GitLab Issues, which features the highest number of operations containing inter-parameter dependency constraints, {\tool} has shown significant improvements in both overall coverage and 2xx coverage, achieving up to a 22.3\% increase. In the services characterized by a high level of operation dependency, it attains a 100\% 2xx coverage for ProShop, outperforming RTG, which achieves  62.5\% coverage. This result validates our design intuition on capturing better dependencies, leading to better coverage.

Several cases correctly identify sequences to test an operation, but the responses of subsequent operations often prove inadequate to support requests for the next operation, especially when responses are empty or there are mismatches between the operation’s implementation and its specification. This issue is prominent in certain operations within GitLab’s services (GitLab Commit, GitLab Repository) or the Bills API, leading to incomplete testing sequences. 


\section{Test generation efficiency (RQ2)}
\begin{table*}[t]
  \centering
  \caption{Test Case Generation Efficiency (RQ2)}
  \label{tab:efficiencyoftestcasegeneration}
  \begin{tabular}{|P{2.2cm}|*{2}{P{0.9cm}|}*{2}{P{0.9cm}|}*{2}{P{0.9cm}|}*{2}{P{0.9cm}|}*{2}{P{0.76cm}|}*{2}{P{0.76cm}|}}
    \hline
    \multicolumn{1}{|c|}{\multirow{2}{*}{REST service}} & \multicolumn{2}{c|}{\makecell{No. test cases \\ generated to cover \\ 2xx SC}} & \multicolumn{2}{c|}{\makecell{No. test cases \\ generated to cover \\ 4xx SC}} & \multicolumn{2}{c|}{\makecell{No. test cases \\ actually covering \\ 2xx SC}} & \multicolumn{2}{c|}{\makecell{No. test cases \\ actually covering \\ 4xx SC}} & \multicolumn{2}{c|}{2xx score (\%)} & \multicolumn{2}{c|}{4xx score (\%)} \\
    \cline{2-13}
    & RTG & KAT & RTG & KAT & RTG & KAT & RTG & KAT & RTG & KAT & RTG & KAT \\
    \hline
    ProShop & 711 & 19 & 569 & 190 & 10 & 16 & 21 & 24 & 1.4 & 84.2 & 3.7 & 12.6 \\
    \hline
    Petstore & 707 & 24 & 956 & 276 & 10 & 15 & 10 & 11 & 1.4 & 62.5 & 1 & 4 \\
    \hline
    Canada Holidays & 6 & 6 & 2 & 2 & 6 & 6 & 2 & 2 & 100 & 100 & 100 & 100 \\
    \hline
    Bills API & 1731 & 132 & 232 & 376 & 14 & 17 & 23 & 26 & 0.8 & 12.9 & 9.9 & 6.9 \\
    \hline
    Genome Nexus & 600 & 36 & - & - & 18 & 23 & - & - & 3 & 63.9 & - & - \\
    \hline
    BingMap-Route & 1413 & 35 & - & - & 4 & 11 & - & - & 0.3 & 31.4 & - & - \\
    \hline
    GitLab-Branch & 704 & 105 & - & - & 5 & 7 & - & - & 0.7 & 6.7 & - & - \\
    \hline
    GitLab-Commit & 4256 & 267 & - & - & 3 & 3 & - & - & 0.1 & 1.1 & - & - \\
    \hline
    GitLab-Groups & 3731 & 650 & - & - & 9 & 10 & - & - & 0.2 & 1.5 & - & - \\
    \hline
    GitLab-Issues & 5674 & 654 & - & - & 11 & 17 & - & - & 0.2 & 2.6 & - & - \\
    \hline
    GitLab-Project & 2008 & 372 & - & - & 21 & 23 & - & - & 1 & 6.2 & - & - \\
    \hline
    GitLab-Repository & 969 & 315 & - & - & 4 & 4 & - & - & 0.4 & 1.3 & - & - \\
     \hline
     Average & 1875.8 & 217.9 & 439.8 & 211 & 9.6 & 12.7 & 14 & 15.8 & 0.5 & 5.8 & 3.2 & 7.5  \\
     \hline
  \end{tabular}
\end{table*}

\subsection{Experimental Methodology}
To answer RQ2, we devised a measure to assess the efficiency of test case generation:

\[\textbf{Efficiency score}_\textbf{R} = \frac{\text{No. TCs actually covering SC in R}}{\text{No. TCs generated to cover SC in R}}\]

with \textit{R}: 2xx or 4xx range.

\indent \textit{No. TCs actually covering SC in R}: the number of test cases that genuinely cover status codes (SC in Table~\ref{tab:dataset_stat}) in the R range. In an operation, we consider its multiple test cases returning the same status code value as one.

\indent \textit{No. TCs generated to cover SC in R}: the number of test cases generated to encompass status codes (SC in Table~\ref{tab:dataset_stat}) in the R range. This variable calculates the minimum number of test cases required to cover SC in the R range, updating the count if new status codes in the same range are covered.

{\em The higher coverage with the lower number of generated test cases implies a higher efficiency score.}

\indent \textbf{Average}: we determine average coverage by dividing the total number of test cases that actually cover status codes within the R range, across all services, by the total number of test cases generated to cover status codes within that range.

\subsection{Experimental Results}

Table \ref{tab:efficiencyoftestcasegeneration} illustrates that the test suite generated by {\tool} comprises fewer instances where the expected coverage is not achieved compared to RTG. Notably, with the ProShop service, KAT successfully covers 84.2\% of the 19 test cases designed for 2xx status codes, while RTG achieves successful coverage for 1.4\% out of the 711 test cases it generates. Moreover, {\tool} shows coverage for 4xx status codes in 12.6\% of the 190 generated test cases, in contrast to RTG's 3.7\% success rate out of the 569 test cases it generates. In general, KAT generates significantly fewer test cases for each coverage type compared to RTG, yet it improves coverage for each service, except for Canada Holidays, where a basically valid value assigned to the required parameter can result in a successful response.

Overall, KAT enhances the efficiency in generating test cases, achieving a 5.3\% improvement for 2xx status codes and a 4.3\% improvement for 4xx status codes. These results affirm that {\tool} enhances the efficiency of test case generation, with the generated test cases considering operation dependencies and aligning request data with parameter constraints.
\section{Failure detection (RQ3)}

\subsection{Experimental Methodology}


\indent \textbf{Errors detection}: the number of errors, which is the count of requests that result in server error status codes (5xx).

\indent \textbf{Undocumented status codes detection}: the number of response status codes that are absent from the OAS files.

\indent \textbf{Mismatches detection}: we rely on trustworthy OAS files. Some previous tools neglected inter-parameter dependencies, leading to false-positive test cases for 2xx status codes. These false-positives occur when the tools listed them as failure cases to test 2xx status codes. Instead, {\tool} leverages these cases to improve the accuracy of testing 4xx status codes, resulting in more reliable coverage for 4xx status codes while allowing precise calculation of 2xx status code coverage.


\subsection{Experimental Results}

\begin{table}[t]
  \centering
  \caption{Failure Detection (RQ3)}
  \label{tab:errorandmismatchfound}
  \begin{tabular}{|P{2.2cm}|*{2}{P{0.5cm}|}*{2}{P{0.65cm}|}*{2}{P{0.55cm}|}}
     \hline
     \multicolumn{1}{|c|}{\multirow{2}{*}{REST service}} & \multicolumn{2}{c|}{500 errors} & \multicolumn{2}{c|}{\makecell{Undocumented \\ status codes}} & \multicolumn{2}{c|}{Mismatches}\\
     \cline{2-7}
      & RTG & KAT & RTG & KAT & RTG & KAT\\
     \hline
     ProShop & 7 & 6  & 0 & 0 & 21 & 14 \\
     \hline
     Petstore & 9 & 8  & 11 & 13 & 15 & 13 \\
     \hline
     Canada Holidays & 0 & 0 & 2 & 2 & 7 & 5 \\
     \hline
     Bills API & 0 & 1  & 0 & 0 & 26 & 26 \\
     \hline
     Genome Nexus  & 0 & 0  & 15 & 14 & 14 & 16\\
     \hline
     BingMap-Route & 4 & 5 & 31 & 32 & 6 & 6\\
     \hline
     GitLab-Branch& 0 & 0 & 15 & 19 & 4 & 5\\
     \hline
     GitLab-Commit & 1 & 0 & 19 & 21 & 3 & 0\\
     \hline
     GitLab-Groups & 1 & 1 & 26 & 28 & 5 & 3 \\
     \hline
     GitLab-Issues & 1 & 1 & 37 & 48 & 8 & 3\\
     \hline
     GitLab-Project & 0 & 1 & 50 & 53 & 8 & 0 \\
     \hline
     GitLab-Repository & 0 & 0 & 15 & 15 & 2 & 3 \\
     \hline
     Total & 23 & 23 & 221 & 245 & 119 & 94\\
     \hline
  \end{tabular}
\end{table}

Table \ref{tab:errorandmismatchfound} shows an equal number of errors detected by both RTG and {\tool}. The variation in error count between each service does not surpass one error. 


In the realm of undocumented status code detection, {\tool} surpasses RTG, detecting 24 more status codes than  RTG. Subsequent investigation underscores the prowess of our approach in identifying previously undocumented status codes, a proficiency that is significantly enhanced when operational dependencies are accurately detected. Consequently, our approach excels at detecting undocumented status codes, encompassing 304, 400, 422 and more. These findings underscore the distinct advantage of our approach in the detection of previously undocumented status codes.

Table \ref{tab:errorandmismatchfound} reveals a notable difference in the average number of mismatches detected between RTG and our approach, {\tool}. RTG detects totally 119 mismatches, whereas {\tool} identifies 94 mismatches. The disparity in mismatch detection can be attributed to several factors. RTG sometimes fails to consider inter-parameter dependency constraints or overlooks fixed parameter values, resulting in the generation of false-positive test cases designed to cover 2xx status codes. Additionally, RTG may incorrectly detect crucial input parameters when subjected to mutation strategies, leading to the creation of false-positive test cases targeting 4xx status codes.

Regarding RQ3, RTG and {\tool} show similar capabilities in detecting errors. KAT adeptly detects status codes absent from the OAS file by considering both operational and inter-parameter dependencies. There is also a notable reduction in false-positive test cases from {\tool}.
\section{Threats to Validity}

There are concerns that may affect the validity of our results.

\vspace{1pt}
{\bfseries Internal validity.} One threat is the hallucination issue of GPT that results in inconsistent outputs when generating ODG, validation scripts, test cases, and test data. We alleviated this threat by using detailed and narrow instructions in prompts and by setting GPT's temperature parameter to zero to control the model's randomness. In addition, generated test cases were run against the working service as a method to validate and reduce the variation in GPT's outputs.

Another threat is concerned with the lack of certainty in~test oracles. By using the real-world services as black-box, we do not know whether the specification or the service implementation is accurate. It is possible that the specification is not up-to-date, or the service has errors. Due to this uncertainty, we reported mismatches in status code between the specification and implementation rather than actual errors found.  

\vspace{2pt}
{\bfseries External validity.} 
Although we evaluated the approaches using real-world applications, they may not represent the general population of API services. We mitigated this by~choosing a diverse set of services, in terms of the application domain, number of operations and parameters, operation dependencies, business rules, and details of operation and parameter descriptions. They are publicly accessible, and many of them were evaluated in previous studies. Thus, the use of these services allows replication and cross-examination in future studies. 
\section{Related work}
Considering various surveys~\cite{kim2022automated,golmohammadi2022testing, ehsan2022restful, martin2022online,  sharma2018automated,marculescu2022faults,martin2021black}, previous approaches to RESTful APIs testing can be classified into two primary directions: black-box testing~\cite{viglianisi2020resttestgen, MartinLopez2021Restest, liu2022morest, atlidakis2019restler,laranjeiro2021black,wu2022combinatorial,corradini2022automated}
and white-box testing~\cite{sahin2021discrete, arcuri2020automated, arcuri2021enhancing,arcuri2020handling, zhang2021adaptive, zhang2021enhancing,zhang2019resource,zhang2021resource}. 

\vspace{2pt}
{\bf Resource dependencies.} Viglianisi {\em et al.} introduced the use of ODG for modeling data dependencies among operations using OpenAPI specifications~\cite{viglianisi2020resttestgen}. Liu {\em et al.} presented the RESTful-service Property Graph (RPG) to provide detailed modeling of producer-consumer dependencies, including schema property equivalence relations. RPG can describe RESTful service behaviors and dynamically update itself with execution feedback~\cite{liu2022morest}. Atlidakis {\em et al.} proposed RESTler to infer dependencies among request types declared in Swagger using a test-generation grammar~\cite{atlidakis2019restler}. Tree-based approaches, like linkage model trees~\cite{stallenberg2021improving} and API trees
~\cite{lin2022forest}, were also considered. In contrast, we focus on constructing the ODG, leveraging natural language descriptions in specifications to guide LLMs for support and revision.



\vspace{2pt}
{\bf Inter-parameter dependency.} 
Martin-Lopez {\em et al.} proposed Inter-parameter Dependency Language (IDL), a domain-specific language for analyzing dependencies among input parameters~\cite{martin2020automated}. RESTest added inter-parameter dependencies to OpenAPI specifications using the IDL~\cite{martin2020automated}, implementing Constraint-Based Testing mode~\cite{MartinLopez2021Restest}. Wu {\em et al.} applied NLP for extracting inter-parameter constraints using pattern-based methods~\cite{wu2022combinatorial}. These approaches formalize inter-parameter dependencies, often requiring manual addition of IDL language or pre-defined NLP patterns, which is time-consuming and potentially error-prone. In contrast, Kim {\em et al.} used Reinforcement Learning to explore dependencies~\cite{kim2023adaptive}, Mirabella {\em et al.} trained a deep-learning model~\cite{mirabella2021deep}, while we leverage LLMs to extract inter-parameter dependencies from OAS files.


\vspace{2pt}
{\bf Natural language processing in API testing.} Kim {\em et al.} introduced NLPtoREST, using NLP techniques to add supplementary OpenAPI rules from the human-readable specification~\cite{kim2023enhancing}. Wanwarang {\em et al.} applied NLP to extract concepts associated with labels, used to query knowledge bases for input values~\cite{wanwarang2020testing}. ARTE explored various API elements, generating realistic test inputs by querying knowledge bases~\cite{alonso2022arte}. Liu {\em et al.} introduced RESTInfer, a two-phase approach inferring parameter constraints from RESTful API descriptions~\cite{liu2022restinfer}.

\section{Conclusion}
KAT is an AI-driven method for automated black-box~testing of RESTful APIs. It uses a large language model and advanced prompting techniques to identify dependencies among operations, input parameters, and between operations and their parameters. KAT incorporates ODG construction in test script creation, detects inter-parameter dependencies, and considers parameter constraints to generate test data on target APIs. 

We evaluated {\tool} using 12 real-world RESTfull API services. The results show that it can automatically generate test cases and data to improve status code coverage, detect more undocumented status codes, and reduces more false positives in these services compared to RTG. It improves an overall status code coverage by 15.7\% over RTG, increasing the coverage from 59.2\% status codes by RTG to 74.9\%. 

\section*{Acknowledgments} Vu Nguyen is partially supported by Vietnam’s NAFOSTED grant NCUD.02-2019.72.
Tien N. Nguyen is supported in part by the US NSF grant CNS-2120386 and
the NSA grant NCAE-C-002-2021.

\balance

\bibliographystyle{IEEEtran}

\bibliography{references}

\begin{thebibliography}{10}
\providecommand{\url}[1]{#1}
\csname url@samestyle\endcsname
\providecommand{\newblock}{\relax}
\providecommand{\bibinfo}[2]{#2}
\providecommand{\BIBentrySTDinterwordspacing}{\spaceskip=0pt\relax}
\providecommand{\BIBentryALTinterwordstretchfactor}{4}
\providecommand{\BIBentryALTinterwordspacing}{\spaceskip=\fontdimen2\font plus
\BIBentryALTinterwordstretchfactor\fontdimen3\font minus \fontdimen4\font\relax}
\providecommand{\BIBforeignlanguage}[2]{{%
\expandafter\ifx\csname l@#1\endcsname\relax
\typeout{** WARNING: IEEEtran.bst: No hyphenation pattern has been}%
\typeout{** loaded for the language `#1'. Using the pattern for}%
\typeout{** the default language instead.}%
\else
\language=\csname l@#1\endcsname
\fi
#2}}
\providecommand{\BIBdecl}{\relax}
\BIBdecl

\bibitem{viglianisi2020resttestgen}
E.~Viglianisi, M.~Dallago, and M.~Ceccato, ``{RestTestGen: automated black-box testing of restful APIs},'' in \emph{2020 IEEE 13th International Conference on Software Testing, Validation and Verification (ICST)}.\hskip 1em plus 0.5em minus 0.4em\relax IEEE, 2020, pp. 142--152.

\bibitem{MartinLopez2021Restest}
A.~Martin-Lopez, S.~Segura, and A.~Ruiz-Cort\'{e}s, ``{RESTest: Automated Black-Box Testing of RESTful Web APIs},'' in \emph{Proceedings of the 30th ACM SIGSOFT International Symposium on Software Testing and Analysis}, ser. ISSTA '21.\hskip 1em plus 0.5em minus 0.4em\relax Association for Computing Machinery, 2021.

\bibitem{arcuri2019restful}
A.~Arcuri, ``Restful api automated test case generation with evomaster,'' \emph{ACM Transactions on Software Engineering and Methodology (TOSEM)}, vol.~28, no.~1, pp. 1--37, 2019.

\bibitem{liu2022morest}
Y.~Liu, Y.~Li, G.~Deng, Y.~Liu, R.~Wan, R.~Wu, D.~Ji, S.~Xu, and M.~Bao, ``Morest: model-based restful api testing with execution feedback,'' in \emph{Proceedings of the 44th International Conference on Software Engineering}, 2022, pp. 1406--1417.

\bibitem{alonso2022arte}
J.~C. Alonso, A.~Martin-Lopez, S.~Segura, J.~M. Garcia, and A.~Ruiz-Cortes, ``{ARTE: Automated Generation of Realistic Test Inputs for Web APIs},'' \emph{IEEE Transactions on Software Engineering}, vol.~49, no.~1, pp. 348--363, 2022.

\bibitem{atlidakis2019restler}
V.~Atlidakis, P.~Godefroid, and M.~Polishchuk, ``{Restler: Stateful rest API fuzzing},'' in \emph{2019 IEEE/ACM 41st International Conference on Software Engineering (ICSE)}.\hskip 1em plus 0.5em minus 0.4em\relax IEEE, 2019, pp. 748--758.

\bibitem{laranjeiro2021black}
N.~Laranjeiro, J.~Agnelo, and J.~Bernardino, ``A black box tool for robustness testing of rest services,'' \emph{IEEE Access}, vol.~9, pp. 24\,738--24\,754, 2021.

\bibitem{brown2020language}
T.~Brown, B.~Mann, N.~Ryder, M.~Subbiah, J.~D. Kaplan, P.~Dhariwal, A.~Neelakantan, P.~Shyam, G.~Sastry, A.~Askell \emph{et~al.}, ``Language models are few-shot learners,'' \emph{Advances in neural information processing systems}, vol.~33, pp. 1877--1901, 2020.

\bibitem{corradini2022resttestgen}
D.~Corradini, A.~Zampieri, M.~Pasqua, and M.~Ceccato, ``{RestTestGen: An Extensible Framework for Automated Black-box Testing of RESTful APIs},'' in \emph{2022 IEEE International Conference on Software Maintenance and Evolution (ICSME)}.\hskip 1em plus 0.5em minus 0.4em\relax IEEE, 2022, pp. 504--508.

\bibitem{apiguru}
\BIBentryALTinterwordspacing
``Apis.guru,'' 2024. [Online]. Available: \url{https://apis.guru/}
\BIBentrySTDinterwordspacing

\bibitem{corradini2021restats}
D.~Corradini, A.~Zampieri, M.~Pasqua, and M.~Ceccato, ``{Restats: A test coverage tool for RESTful APIs},'' in \emph{2021 IEEE International Conference on Software Maintenance and Evolution (ICSME)}.\hskip 1em plus 0.5em minus 0.4em\relax IEEE, 2021, pp. 594--598.

\bibitem{vassiliou2020simple}
T.~Vassiliou-Gioles, ``A simple, lightweight framework for testing restful services with ttcn-3,'' in \emph{2020 IEEE 20th International Conference on Software Quality, Reliability and Security Companion (QRS-C)}.\hskip 1em plus 0.5em minus 0.4em\relax IEEE, 2020, pp. 498--505.

\bibitem{kim2022automated}
M.~Kim, Q.~Xin, S.~Sinha, and A.~Orso, ``Automated test generation for rest apis: No time to rest yet,'' in \emph{Proceedings of the 31st ACM SIGSOFT International Symposium on Software Testing and Analysis}, 2022, pp. 289--301.

\bibitem{kim2023adaptive}
M.~Kim, S.~Sinha, and A.~Orso, ``{Adaptive REST API Testing with Reinforcement Learning},'' \emph{arXiv preprint arXiv:2309.04583}, 2023.

\bibitem{kim2023enhancing}
M.~Kim, D.~Corradini, S.~Sinha, A.~Orso, M.~Pasqua, R.~Tzoref-Brill, and M.~Ceccato, ``{Enhancing REST API Testing with NLP Techniques},'' in \emph{Proceedings of the 32nd ACM SIGSOFT International Symposium on Software Testing and Analysis}, 2023, pp. 1232--1243.

\bibitem{wu2022combinatorial}
H.~Wu, L.~Xu, X.~Niu, and C.~Nie, ``{Combinatorial testing of restful APIs},'' in \emph{Proceedings of the 44th International Conference on Software Engineering}, 2022, pp. 426--437.

\bibitem{yamamoto2021efficient}
K.~Yamamoto, ``{Efficient penetration of API sequences to test stateful RESTful services},'' in \emph{2021 IEEE International Conference on Web Services (ICWS)}.\hskip 1em plus 0.5em minus 0.4em\relax IEEE, 2021, pp. 734--740.

\bibitem{karlsson2020quickrest}
S.~Karlsson, A.~{\v{C}}au{\v{s}}evi{\'c}, and D.~Sundmark, ``{QuickREST: Property-based test generation of OpenAPI-described RESTful APIs},'' in \emph{2020 IEEE 13th International Conference on Software Testing, Validation and Verification (ICST)}.\hskip 1em plus 0.5em minus 0.4em\relax IEEE, 2020, pp. 131--141.

\bibitem{lin2022forest}
J.~Lin, T.~Li, Y.~Chen, G.~Wei, J.~Lin, S.~Zhang, and H.~Xu, ``{foREST: A Tree-based Approach for Fuzzing RESTful APIs},'' \emph{arXiv preprint arXiv:2203.02906}, 2022.

\bibitem{proshop}
\BIBentryALTinterwordspacing
``{Proshop API},'' 2024. [Online]. Available: \url{https://anonymous.4open.science/r/proshop-api-B648}
\BIBentrySTDinterwordspacing

\bibitem{golmohammadi2022testing}
A.~Golmohammadi, M.~Zhang, and A.~Arcuri, ``{Testing RESTful APIs: A Survey},'' \emph{ACM Transactions on Software Engineering and Methodology}, 2022.

\bibitem{ehsan2022restful}
A.~Ehsan, M.~A.~M. Abuhaliqa, C.~Catal, and D.~Mishra, ``{RESTful API testing methodologies: Rationale, challenges, and solution directions},'' \emph{Applied Sciences}, vol.~12, no.~9, p. 4369, 2022.

\bibitem{martin2022online}
A.~Martin-Lopez, S.~Segura, and A.~Ruiz-Cort{\'e}s, ``{Online testing of RESTful APIs: Promises and challenges},'' in \emph{Proceedings of the 30th ACM Joint European Software Engineering Conference and Symposium on the Foundations of Software Engineering}, 2022, pp. 408--420.

\bibitem{sharma2018automated}
A.~Sharma, M.~Revathi \emph{et~al.}, ``{Automated API testing},'' in \emph{2018 3rd International Conference on Inventive Computation Technologies (ICICT)}.\hskip 1em plus 0.5em minus 0.4em\relax IEEE, 2018, pp. 788--791.

\bibitem{marculescu2022faults}
B.~Marculescu, M.~Zhang, and A.~Arcuri, ``{On the faults found in rest APIs by automated test generation},'' \emph{ACM Transactions on Software Engineering and Methodology (TOSEM)}, vol.~31, no.~3, pp. 1--43, 2022.

\bibitem{martin2021black}
A.~Martin-Lopez, A.~Arcuri, S.~Segura, and A.~Ruiz-Cort{\'e}s, ``{Black-box and white-box test case generation for RESTful APIs: Enemies or allies?}'' in \emph{2021 IEEE 32nd International Symposium on Software Reliability Engineering (ISSRE)}.\hskip 1em plus 0.5em minus 0.4em\relax IEEE, 2021, pp. 231--241.

\bibitem{corradini2022automated}
D.~Corradini, A.~Zampieri, M.~Pasqua, E.~Viglianisi, M.~Dallago, and M.~Ceccato, ``{Automated black-box testing of nominal and error scenarios in RESTful APIs},'' \emph{Software Testing, Verification and Reliability}, vol.~32, no.~5, p. e1808, 2022.

\bibitem{sahin2021discrete}
O.~Sahin and B.~Akay, ``{A discrete dynamic artificial bee colony with hyper-scout for RESTful web service API test suite generation},'' \emph{Applied Soft Computing}, vol. 104, p. 107246, 2021.

\bibitem{arcuri2020automated}
A.~Arcuri, ``{Automated black-and white-box testing of restful APIs with evomaster},'' \emph{IEEE Software}, vol.~38, no.~3, pp. 72--78, 2020.

\bibitem{arcuri2021enhancing}
A.~Arcuri and J.~P. Galeotti, ``{Enhancing search-based testing with testability transformations for existing APIs},'' \emph{ACM Transactions on Software Engineering and Methodology (TOSEM)}, vol.~31, no.~1, pp. 1--34, 2021.

\bibitem{arcuri2020handling}
------, ``Handling sql databases in automated system test generation,'' \emph{ACM Transactions on Software Engineering and Methodology (TOSEM)}, vol.~29, no.~4, pp. 1--31, 2020.

\bibitem{zhang2021adaptive}
M.~Zhang and A.~Arcuri, ``{Adaptive hypermutation for search-based system test generation: A study on rest APIs with EvoMaster},'' \emph{ACM Transactions on Software Engineering and Methodology (TOSEM)}, vol.~31, no.~1, pp. 1--52, 2021.

\bibitem{zhang2021enhancing}
------, ``{Enhancing resource-based test case generation for RESTful APIs with SQL handling},'' in \emph{International Symposium on Search Based Software Engineering}.\hskip 1em plus 0.5em minus 0.4em\relax Springer, 2021, pp. 103--117.

\bibitem{zhang2019resource}
M.~Zhang, B.~Marculescu, and A.~Arcuri, ``{Resource-based test case generation for RESTful web services},'' in \emph{Proceedings of the genetic and evolutionary computation conference}, 2019, pp. 1426--1434.

\bibitem{zhang2021resource}
------, ``Resource and dependency based test case generation for restful web services,'' \emph{Empirical Software Engineering}, vol.~26, no.~4, p.~76, 2021.

\bibitem{stallenberg2021improving}
D.~Stallenberg, M.~Olsthoorn, and A.~Panichella, ``{Improving test case generation for rest APIs through hierarchical clustering},'' in \emph{2021 36th IEEE/ACM International Conference on Automated Software Engineering (ASE)}.\hskip 1em plus 0.5em minus 0.4em\relax IEEE, 2021, pp. 117--128.

\bibitem{martin2020automated}
A.~Martin-Lopez, ``{Automated analysis of inter-parameter dependencies in web APIs},'' in \emph{Proceedings of the ACM/IEEE 42nd International Conference on Software Engineering: Companion Proceedings}, 2020, pp. 140--142.

\bibitem{mirabella2021deep}
A.~G. Mirabella, A.~Martin-Lopez, S.~Segura, L.~Valencia-Cabrera, and A.~Ruiz-Cort{\'e}s, ``{Deep learning-based prediction of test input validity for restful APIs},'' in \emph{2021 IEEE/ACM Third International Workshop on Deep Learning for Testing and Testing for Deep Learning (DeepTest)}.\hskip 1em plus 0.5em minus 0.4em\relax IEEE, 2021, pp. 9--16.

\bibitem{wanwarang2020testing}
T.~Wanwarang, N.~P. Borges~Jr, L.~Bettscheider, and A.~Zeller, ``Testing apps with real-world inputs,'' in \emph{Proceedings of the IEEE/ACM 1st International Conference on Automation of Software Test}, 2020, pp. 1--10.

\bibitem{liu2022restinfer}
Y.~Liu, ``{RESTInfer: automated inferring parameter constraints from natural language RESTful API descriptions},'' in \emph{Proceedings of the 30th ACM Joint European Software Engineering Conference and Symposium on the Foundations of Software Engineering}, 2022, pp. 1816--1818.

\end{thebibliography}

\end{document}